\documentclass[pra,aps,twocolumn,floatfix, 10pt]{revtex4-2}
\usepackage{graphicx}
\graphicspath{ {./images/} }
\usepackage{subfig}
\usepackage{xcolor}

\usepackage{natbib}
\usepackage{amssymb}

\usepackage{xcolor}
\usepackage{amsmath}
\usepackage{soul}
\usepackage[bookmarks=false, hidelinks]{hyperref}
\usepackage[hidelinks]{hyperref}

\begin{document}
\title{An Alternative Approach to the Osmotic Second Virial Coefficient of Protein Solutions and its Application to Liquid-Liquid Phase Separation}

\author{Furio Surfaro}
\email{furio.surfaro@uni-tuebingen.de}
\author{Ralph Maier, Kai-Florian Pastryk, Fajun Zhang}
\author{Frank Schreiber}
\affiliation{Institute of Applied Physics, University of T{\"u}bingen, 72076 T{\"u}bingen, Germany}
\author{Roland Roth}
\email{roland.roth@uni-tuebingen.de}
\affiliation{Institute of Theoretical Physics, University of T{\"u}bingen, 72076 T{\"u}bingen, Germany}

\begin{abstract}
The osmotic second virial coefficient $B_2$ is an important parameter to describe the interactions and phase behaviour of protein solutions, including colloidal systems and macromolecular solutions. Another key parameter to describe the driving force of the nucleation of a new phase is the supersaturation, which is used in the Classical Nucleation Theory (CNT) framework and is connected with the favorable contribution in the Gibbs free energy in the bulk solution. In this article we establish a connection between $B_2$ calculated from small angle X-ray scattering (SAXS) data with the values of $B_2$ obtained from supersaturation measurements using thermodynamics considerations. The values of the second virial coefficient calculated employing this method agree with those determined via SAXS in the region near the liquid-liquid phase separation (LLPS) border for human serum albumin (HSA) and bovine serum albumin (BSA). 
The general relations adopted are shown to be useful for the estimation of the second virial coefficient $B_2$ for globular proteins, in proximity of the binoidal biphasic coexistent region. 
\end{abstract}

\maketitle

\section{Introduction}

Liquid-liquid phase separation (LLPS) in protein solutions and
protein crystallization, as well as in different macromolecular and colloidal systems, are interesting and important phenomena 
that have far-reaching consequences in physics, chemistry, biology
and medicine \cite{alberti2019considerations,zbinden2020phase, you2020phasepdb,zhang2012role,hyman2014liquid,zhang2020liquid}. LLPS may serve as a mechanism for cellular organization \cite{hyman2014liquid}, or as a precursor on the pathway towards crystallization \cite{maier2020protein,maier2021protein}.
Despite the relatively simple observation of LLPS and protein
crystals on the macroscopic scale, the rich phenomenology of protein
solutions and their microscopic behavior are not fully understood yet.
Given the complex nature of protein molecules and interactions on the
microscopic scale, usually the key parameters used to describe
protein solutions and their behavior are calculated with respect to
macroscopic properties that are more easily accessible. Thus,
thermodynamic relations are applied to provide insights about the
driving force of the nucleation process of a new phase. 

An important parameter, often used to describe the tendency of a
protein solution to phase separate, is the second virial coefficient
$B_2$, that measures the overall strength of the effective 
protein-protein interaction U({\bf r},$\Omega)$, that depends on
the relative position {\bf r} and orientation $\Omega$ between
the proteins, averaged over all possible distances and orienations
via 
\begin{equation} \label{eq:b2def}
B_2(T) = -\frac{1}{2} \int [e^{-U(r,\Omega)/kT} -1] d\Omega \, d^3r.
\end{equation}
Since the effective interaction potential between proteins 
is typically not known,
a direct calculation of the second virial coefficient with
Eq.~(\ref{eq:b2def}) is impossible. Still, $B_2$ is accessible either
by thermodynamic considerations, or experimentally. If $B_2$ is 
positive, the effective interaction is overall repulsive, while it
is attractive for negative values of $B_2$. For LLPS a sufficiently
strong attraction is required and it has been observed for several
colloidal systems $B_2/B_{HS} \approx$ -1.5 
\cite{doi:10.1063/1.481106} close to a critical point and more
negative in the LLPS region. For protein solutions that display
LLPS, even though the effective interaction is much more 
complicated than in the colloidal system, similar observations
have been made close to the LLPS region 
\cite{wolf2014effective, platten2015extended}. Furthermore,
it seems that frequently $B_2$ falls into a narrow range when crystallization
occurs, called crystallization slot \cite{george1994predicting}.

In order to obtain information about $B_2$, different experimental
techniques are used, such as equilibrium sedimentation, dynamic 
light scattering and small angle x-ray scattering (SAXS). However,
these different approaches do not always lead to the same values
and the estimated results cannot easily be compared directly \cite{winzor2007nonequivalence, deszczynski2006negative}.
Additionally, the estimation of the $B_2$ value requires different
approximations due to the complex nature of the interactions.

Eq.~(\ref{eq:b2def}) is applied for the calculation of $B_2$, assuming different forms of simplified
interaction potentials $U(r)$, that typically only depend on the
relative distance {\bf r}, in order to improve the quality of the fit, i.e., assuming isotropy
\cite{zhang2012charge, narayanan2003protein,zhang2007protein, de1989adhesive}.

In this work, we establish a thermodynamic relation in order to
connect the osmotic second virial coefficient $B_2$ with the 
supersaturation. Our method, explained in Sec.~\ref{sec:th},
can be applied if the system, such as a protein solution, shows a
meta-stable LLPS, and relies solely on easily accessible macroscopic 
observables. We explain different experimental procedures that have 
successfully been used for the determination of $B_2$ so far in 
Sec.~\ref{sec:ex}. In Sec.~\ref{sec:res}, we compare results from our 
approach to those from SAXS fitting for human serum albumin (HSA) and 
bovine serum albumin (BSA) in solution with trivalent salt ions, which were shown to be a versatile method to manipulate interactions \cite{Matsarskaia_2018_PhysChemChemPhys,Zhang_2008_PhysRevLett}, before we conclude in Sec.~\ref{sec:con}. 

\section{Theory} \label{sec:th}
We start by considering a protein solution with concentration $c$. Its
chemical potential $\mu$, relative to a dilute reference system with
concentration $c_0$ and chemical potential $\mu_0$, is given by
\begin{equation} \label{eq:mua}
\mu = \mu_0 + RT \ln\left(\frac{\gamma c}{c_0}\right),
\end{equation}
where $\gamma$ is the activity coefficient that takes the deviations of 
the solution from ideal behaviour into account. Here, we assume that all
effects of the solvent, ions and crowding are included by $\gamma$.
By definition, $\gamma$ is equal to 1 for ideal solutions, i.e., at low protein
concentrations. If $\gamma$ deviates from unity, it indicates a 
concentration that is sufficiently high so that particles interact. A 
value of $\gamma$ smaller than 1 corresponds to overall interactions 
between particles that are attractive, while a value larger than 1 hints 
towards overall interactions that are repulsive. 
We wish to describe an experimental situation, where a protein solution
is prepared at an initial concentration $c_i$, that lies within the LLPS
region, so that the solution phase separates into a low density phase with
concentration $c_l$ and a high density phase with concentration $c_h$.
For protein solutions the LLPS is usually meta-stable, which makes the
determination of $c_l$ and $c_h$ far from trivial.
However, we know that along the tie line of the phase separation one can observe chemical and
mechanical equilibrium, i.e., constant chemical potential and constant
osmotic pressure. In the following, we will make use of the chemical
equilibrium: $\mu(c_i)=\mu(c_l)=\mu(c_h)$. A schematic phase diagram describing this process can be found in reference \cite{zhang2021nonclassical}.
After the LLPS occurs one can observe that the concentration of the low density
phase reduces to its equilibrium value $c_{eq}$, as the solubility line is
reached. During this step the high density phase is totally or partially consumed, if additional phases are present, i.e. precipitates or crystals. We assume that the
resulting solution at concentration $c_{eq}$ is sufficiently dilute that
we can treat it as an ideal solution with $\gamma_{eq}=1$. The change in 
the chemical potential during this process is given by
$\Delta \mu = \mu(c_{eq})-\mu(c_l) = \mu(c_{eq})-\mu(c_i)$, where we
have employed the assumption that the chemical potential of the low density phase
equals that of the initial solution (chemical equilibrium along the tie
line of the LLPS). With the chemical potential given in Eq.~(\ref{eq:mua})
we can express change in the chemical potential as
\begin{equation} \label{eq:dmua}
    \Delta \mu_a = R T \ln\left( \frac{c_{eq}}{\gamma_i c_i}\right),
\end{equation}
where the index $a$ in Eq.~(\ref{eq:dmua}) indicates that we employ the
{\em activity} formula of the chemical potential. Note that in this
expression, $c_i$ and $c_{eq}$ are experimentally accessible, while $\gamma_i$
is unknown.

Following McMillan and Mayer \cite{mcmillan1945statistical}, we can also treat
our system as an effective two-component system of a solvent, including
ions, which we denote as component 1, and the proteins, which is component 2.
The chemical potential of component 1, $\mu_1$, can be expanded into a virial
series in the concentration $c$ of the proteins to obtain
\begin{equation}	
\mu_1 = \mu_1^\ominus - RT V_1 c_{{2}} [\frac{1}{M} + B_{2} c_{{2}} +B_3 c_{{2}}^{2} + \dots],
\end{equation}
where $V_1$ is the molar volume of the solvent, and $M$ the molecular 
weight of the protein. $B_2$, $B_3$, $\dots$ are the second, third, and 
higher virial coefficients. If the protein solution is sufficiently 
dilute, as we have assumed in the low density phase, it is possible to 
truncate the virial expansion at the second term
\citep{guo1999correlation}.
By differentiating $\mu_1$ w.r.t. the protein concentration
\begin{equation} \label{eq:mm}
\left(\frac{d \mu_1}{d c_2}\right)_{p,T} = -{\frac{RTV_{m,1}}{M} [1+2 B_2 M c_2}],
\end{equation}
and employing the Gibbs-Duhem relation \cite{duhem1898general}, it is possible to connect the 
chemical potential of the solvent with the chemical potential of the proteins 
to obtain
\begin{equation} \label{eq:gd}
 \left(\frac{d \mu_{2}}{d c_{2}}\right)_{p,T}= -{\frac{M}{c_{{2}} V_{m,1}}} \left(\frac{d \mu_1}{d c_{2}}\right)_{p,T}.
 \end{equation}

Combining Eqs.(\ref{eq:mm}) and (\ref{eq:gd}), we can calculate the 
chemical potential, relative to a low density reference state with concentration
$c_0$ as
\begin{equation} \label{eq:mumm}
    \mu(c) = \mu_0 + R T \ln \left(\frac{c}{c_0}\right)+2 R T B_2 M (c-c_0),
\end{equation}
which is an alternative to Eq.~(\ref{eq:mua}), as long as the virial expansion
can be truncated after the second term. We are now in the position to express
the change in the chemical potential from the initial solution at concentration
$c_i$ to its final state with concentration $c_{eq}$ by
\begin{equation} \label{eq:dmumm}
    \Delta \mu_{MM} = R T \ln\left(\frac{c_{eq}}{c_i}\right) + 2RT B_2 M (c_{eq}-c_i),
\end{equation}
which is an alternative expression to Eq.~(\ref{eq:dmua}), where $\Delta\mu_{MM}$ indicate the chemical potential obtained from the McMillan-Mayer theory. By demanding that
$\Delta \mu_a = \Delta \mu_{MM}$, we find that the second virial coefficient
takes the form,
\begin{equation}
    B_2 = - \frac{\ln(\gamma_i)}{2 M (c_{eq}-c_i)},
\end{equation}
which still contains the unknown activity coefficient $\gamma_i$.

In order to express the second virial coefficient solely by quantities that
are accessible in the experiment we consider again the chemical
equilibrium between the initial solution and the low density phase along the
tie line, $\mu(c_i)=\mu(c_l)$, from which the following equation is obtained
\begin{equation}
    \ln(\gamma_i) = \ln\left(\frac{\gamma_l c_l}{c_i}\right) \approx \ln\left(\frac{\gamma_{eq} c_{eq}}{c_i}\right) = \ln\left(\frac{c_{eq}}{c_i}\right),
\end{equation}
where we have approximated the activity of the low density phase, which is
not directly accessible, by that of the equilibrium solution, for which we
have assumed that its activity coefficient $\gamma_{eq}=1$. With this
assumption we reach our theoretical main result, an estimate of
the second virial coefficient based on quantities that are easily accessible:
\begin{equation} \label{eq:b2}
    B_2 = - \frac{\ln\left(\frac{c_{eq}}{c_i}\right)}{2 M (c_{eq}-c_i)}.
\end{equation}
Note that assuming an interaction potential with a fixed interaction distance, like in the SAXS fitting procedure, or using a linear interpolation of the light scattering intensity like in the DLS framework, the second virial coefficient, should not depend on $c_i$, despite that, dependence on initial concentration was already found in systems of cyclodextrins using a virial expansion with respect to the structure factor that follow the same concentration dependence described by the Eq.~(\ref{eq:b2}) \cite{kusmin2008native, kusmin2007hydration}.In order to remove the concentration dependence two assumptions can be adopted. The first one is to average the results over the concentration interval explored, keeping the salt-protein ratio constant, the second one is to normalize the values with respect to the first concentration point at any given initial concentration where LLPS is showed, as described in Sec.IV D.
The second virial coefficient obtained is normalized by the second virial coefficient of hard spheres \cite{schreiber2011virial}. In order to demonstrate the validity of Eq.~(\ref{eq:b2}) we apply it to different protein solutions that display LLPS, induced by trivalent salts.

\section{Experiments and Methods} \label{sec:ex}
\subsection{Materials and sample preparation}
Proteins and salt were purchased from Merck, and used as received. 
The purities were 98\% for BSA (product no. A7906), 97\% for HSA (product no. A9511) and 99.99\% for CeCl3 (product no. 429406). Stock solutions were prepared by dissolving the protein and salt in deionized (18.2 M$\Omega$), degassed Millipore water. The resulting concentration of protein was determined with an ultraviolet visible (UV-vis) spectrophotometer (Cary 50 UV-vis spectrometer, Varian Technologies) using an extinction coefficient of 0.667 mL mg$^{-1}$ cm$^{-1}$ for BSA, and 0.531 mL mg$^{-1}$ cm$^{-1}$ for HSA at a wavelength of 278 nm \cite{fasman2018crc}. All samples were prepared by mixing deionized, degassed Millipore water, protein stock solution and salt stock solution. All samples had a pH (between 6.3
and 7.0) above the respective pI of HSA and BSA, measured with a pH-Meter from Mettler Toledo (Germany). No buffer was added as neutral trivalent salts (i.e, CeCl$_3$) do not induce significant pH variation. All samples were prepared, stored and investigated at 21 ± 1 °C. When the LLPS is approached in the phase diagram,  Fig.1, the solution phase-separates into a dense, yellowish phase and in a low density, dilute phase. In addition, for HSA, crystallization was also observed \cite{buchholz2023kinetics}, during this process the dense phase is almost totally consumed \cite{maier2021protein}. After 14 days, the concentration of the resulting dilute equilibrium phase was determined via UV-vis spectroscopy.
\subsection{Salt versus protein concentration phase diagram}
Samples of protein concentrations ($c_p$) at  35, 50, 65, 80, and 100 mg/
mL were prepared for BSA and HSA, varying salt concentrations ($c_s$). The mean value of $c_s$ of the last clear and first turbid sample is referred to as $c^{\star}$ and the last turbid and first clear sample as $c^{\star \star}$. The macroscopic phase separation was ensured by visual inspection.
\begin{figure}
\includegraphics[scale= 0.45]{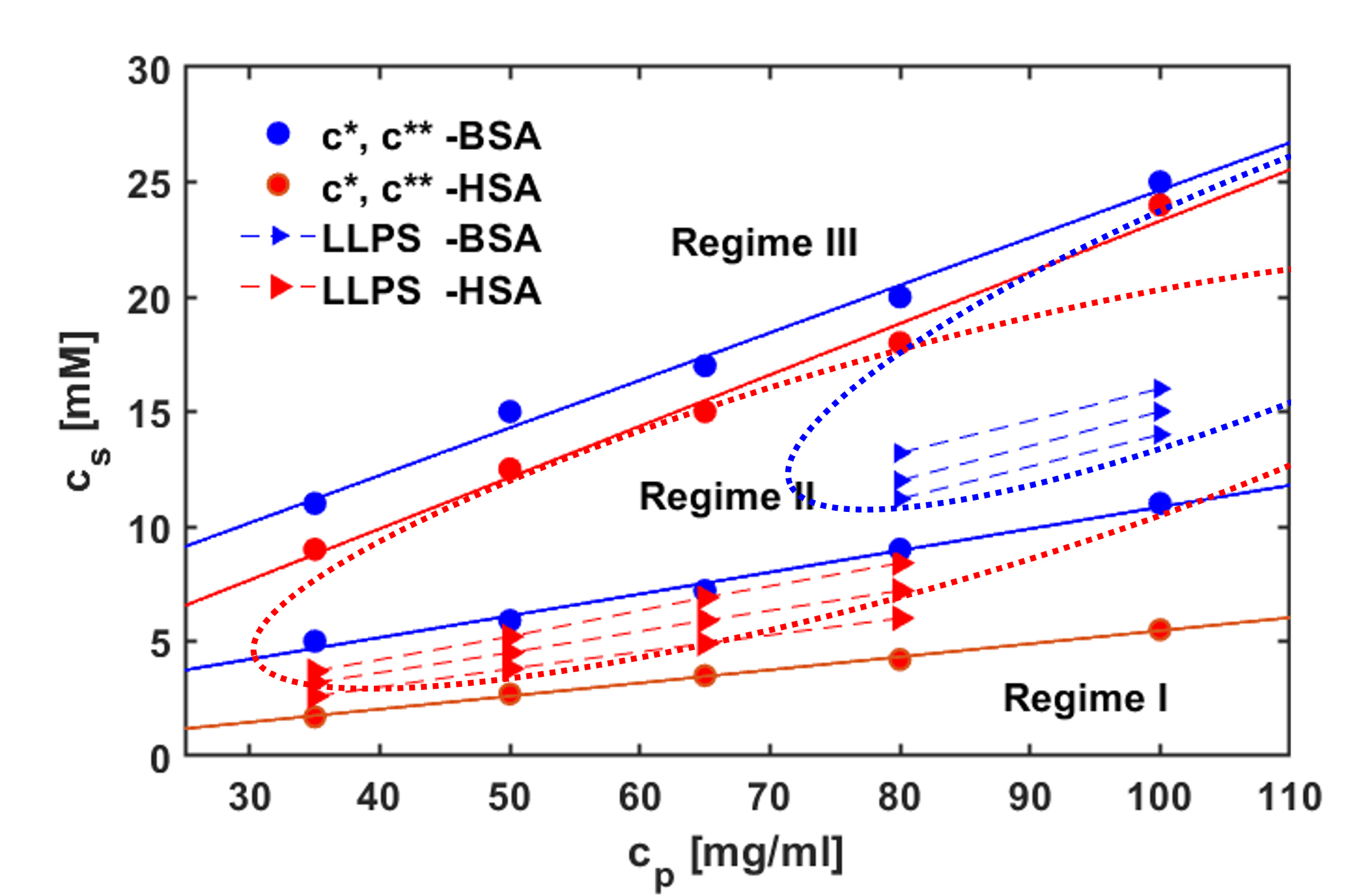}
\caption{\label{fig:phase} The experimental phase diagram displays triangular points that indicate the conditions where the samples inside the LLPS were prepared. Points that share the same $c_s$ over $c_p$ ratio are connected by dashed lines. The phase diagram boundaries are marked by full dots connected by solid lines. The $c^*$ border is indicated by lower solid lines, while the upper solid lines represent the $c^{**}$ border. Moreover, the phase diagram includes red and blue elliptical regions that represent a hypothetical complete LLPS loop for HSA and BSA} \cite{maier2021human}.

\end{figure}
No additional investigation with respect to the distribution of the salt in the two phases was made, and the concentration of the high density phase was not determined. Therefore, the points depicted in the phase diagram refer to the preparation conditions. For the complete knowledge of the LLPS loop additional analysis is needed, such as those performed in the references \cite{wolf2014effective, matsarskaia2016cation, zhang2012charge}.
The phase diagram depicted in Fig.~\ref{fig:phase} shows that for BSA the amount of salt required to reach the $c^{\star}$ border is higher than for HSA. Such behavior could be due to the different surface charge of the proteins at such condition. BSA is slightly more negative with an overall charge of -11 compared to the value of HSA, which is around -9 \cite{maier2021human}. Therefore, more salt has to be added to neutralize the excess negatively charged residues with counterions and trigger the aggregation. Also, the different locations of the $c^{\star \star}$ borders could be explained as a consequence of an increased amount of counterions in BSA compared to HSA to trigger the pair repulsion (reentrant condensation).

\subsection{Small-Angle X-ray Scattering}
SAXS measurements were performed at the Petra III beamline P12 (Hamburg, Germany).
All the details related to the experimental procedures and the data analysis are available in reference \cite{maier2021human}.

For the determination of the reduced osmotic second virial coefficient $B'_2$ = $B_2$ \slash$B_{HS}$
the data were fitted with a sticky hard sphere (SHS) model, provided in the NIST macros, using IGOR Pro 6.37.

In this model the potential of mean force between proteins is assumed to be:
\begin{equation}
\beta U(r) = \begin{cases} 
\infty &  \text{for } r < \sigma, \\ 
-\beta U_o = \ln\left(\frac{12\tau\Delta}{\sigma + \Delta}\right)  &  \text{for } \sigma < r < \sigma + \Delta, \\ 
0 & \text{for } r > \sigma + \Delta.
\end{cases}
\end{equation}

where $\tau$ is the stickiness parameter, $\beta$ the inverse temperature and $\Delta$ the width of the square well. 
A perturbative solution of the Percus-Yevick closure relation was used to calculate the structure factor \cite{baxter1968percus}.
The analysis was performed using HSA and BSA solution with an initial concentration of 50 mg$\slash$ml.
For BSA, no LLPS or crystallization were observed at such concentration, HSA instead displays a rich phase behavior with both crystallization and LLPS. These differences were already investigated in our previous studies \cite{maier2021human}.
The normalized second virial coefficient is thus calculated in the limit of $\Delta\rightarrow 0$  using:
\begin{equation}
\lim_{\Delta\to\ 0} \frac{B_2}{B_{HS}} = 1-\frac{1}{4\tau} \ .
\end{equation}
\section{Results and Discussion} \label{sec:res}
\subsection{Discussion of salt versus protein concentration phase diagram}
The determination of the phase diagram was carried out by visual inspection.
For small amounts of salt, the overall interactions between proteins are repulsive, resulting in a clear solution without macroscopic aggregation (Regime I). Increasing the salt concentration the system reaches the boundary called c$^\star$ where the last clear sample and the first turbid was observed (Regime II).
Here, the interactions between proteins are mostly attractive due to the ion bridging effect \cite{roosen2014ion}.
Further addition of trivalent salt (CeCl$_3$) moves the system towards the second boundary c$^{\star\star}$ where the resulting solution becomes clear again (Regime III or reentrant). A possible explanation of this behavior is the charge inversion of the proteins due to the complete saturation of the binding sites with trivalent ions, leading to strong repulsive interactions \cite{roosen2014ion}.
The experimental data points are fitted using a linear regression.
It is interesting to note that the critical threshold value of protein concentration for macroscopic LLPS is shifted by about two-three times if we compare HSA and BSA (35-40 mg$\slash$ml and 72-80 mg$\slash$ml) for metastable LLPS \cite{maier2020protein, maier2021human}. This behavior reflects the trend observed in the calculation of the osmotic second virial coefficient from supersaturation measurements explained in the next chapter,  where it is observed that the values of the second virial coefficient for BSA is 2 to 3 times bigger than the values of the HSA system, in agreement with the magnitude of the attractive interactions found in this work. If the interactions are not sufficiently attractive at low concentration of protein, an increasing protein concentration greatly shortens the distance between proteins enhancing the likelihood of formation of a critical cluster large enough to allow the nucleation of a new phase.
An additional threshold in concentration in Regime II is reached at high protein concentration, where the system does not undergo phase separation. This might be due to additional short range repulsion that can affect the overall attractive interactions in crowded environments as well as different entropic contribution \cite{zhang2007protein}.
Additional investigations are needed to understand protein interactions in Regime II at higher protein volume fraction, where additional phenomena are observed for several systems, such as kinetic aggregation and amorphous precipitation or gelation \cite{dessau2011protein, de1993aggregation}.

\subsection{Real time observation of phase separation}
The phase separation was followed by visual inspection in glass vials and on glass slides under the microscope. The general behavior observed reveals that close to the binodal, the solutions form a stable high density phase, in the case of BSA, and a meta-stable high density phase, in the case of HSA. This is connected with a different crystallization behavior of the two proteins. In fact, as shown in our previous work a more stable crystalline phase exist for HSA at such condition while for BSA, crystallization does not take place.
This is due to the different intermolecular interactions between this two proteins, since the BSA is more hydrophilic, additional hydrophobic interactions necessary for crystallization are suppressed. Therefore, for the HSA solution the LLPS is formed as kinetically driven process upon salt addition followed by the formation of the crystals from the low density phase that is the thermodynamic stable phase as here shown:
\begin{center} 
$HSA(dense)\leftrightharpoons HSA(dilute)\rightarrow HSA(xtal).$
\end{center}
The system first phase separates in a high density and low density phase until a dynamic equilibrium is reached between them.
They remain stable until the first crystals appear in the low density phase.
Therefore, the variation of the osmotic pressure between the two previously coexistent phases leads to the consumption of the dense liquid droplets to preserve the initial equilibrium. However, if the driving force of the crystallization process is sufficiently strong, an additional three phases equilibrium cannot be reached and the process continues until the complete high density phase is consumed and the equilibrium concentration $c_{eq}$ is reached, which corresponds to the solubility limit \cite{maier2021human}.

The growth of the crystals generally is completed between 7 and 10 days at these conditions. In order to ensure that the phase conversion was completed, a period of 14 days from the sample preparation was chosen for our supersaturation measurements.
\subsection{Calculation of the second virial coefficient from the supersaturation}
\begin{figure}
\begin{center}
\includegraphics[scale=0.5]{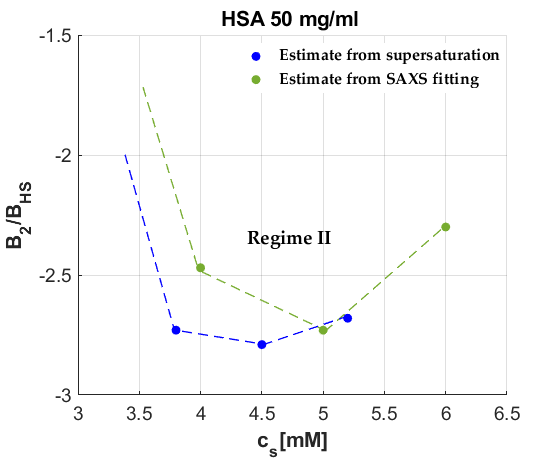}  
\caption{\label{fig:b2hsa} Normalized second virial coefficient estimate from supersaturation and SAXS fitting procedure, in the SAXS data points, the errors bars are smaller then the symbols shown here. For the points obtained by supersaturation measurements the error is estimate to be around 10$\%$ -- largely due to the uncertainty of the concentration measurements, determined via UV-vis spectroscopy at equilibrium. The dotted lines act as a guide for the eyes. Although the different routes to $B_2$ make use of different data, the agreement between the estimates is good.}
\end{center}
\end{figure}

Here, the experimental setup for the calculation of the second virial coefficient and the results are discussed. Upon the addition of CeCl$_3$ the samples in glass vials, sealed with parafilm, were kept at temperature of 21 ± 1 °C for 14 days. This was done to ensure complete phase separation and crystallization. If HSA samples after preparation are centrifuged at 5000 rpm for 10 min and the high density phase is precipitated, the resulting concentration of the low density phase is always higher than the equilibrium concentration obtained 14 days after preparation. This behavior is essentially due to the crystallization that affects the final equilibrium concentration. For the BSA samples, which only show phase separation but no crystallization, the equilibrium concentration after 14 days is close to the values obtained after centrifuging.
Therefore, to consider the total contribution (phase separation and crystallization) on the partitioning, only the values of the concentration obtained after 14 days are employed for the estimation of the osmotic second virial coefficient. Considering the normalized second virial coefficient obtained as a ratio of  Eq.~(\ref{eq:b2}) and $B_{HS}$ in ml$\cdot$mol$\cdot$g$^{-2}$, the only  parameter not known experimentally is the hydrodynamic hard sphere radius.
Hence, for the calculation of the hydrodynamic radius a previously resolved crystallographic structure in presence of YCl$_3$ was used as an input in the software HullRad from fluidic Analytics, which  uses a convex hull model to estimate the hydrodynamic volume of a macromolecule \cite{fleming2018hullrad}.
Further details about this method are given in references \cite{fleming2018hullrad, barber1996quickhull}.
The resulting value is estimated to $R_h$ = 3.55 nm and was used together with a molecular weight of 66.4 kDa for the calculation of the normalized second virial coefficient of both HSA and BSA from the supersaturation measurements.
A similar value for $R_h$ was found experimentally using SAXS on BSA diluite solutions in reference \cite{roosen2011protein}, where the scattering curves are well fitted by a prolate ellipsoid with an effective hydrodynamic radii of $R_h$ = 3.62 nm. In the same reference a value of $R_h$ = 3.66 nm obtained via DLS is also found. For the calculation of the second virial coefficient from the SAXS fitting procedure, an ellipsoid form factor was chosen. The ellipsoid form factor chosen was the same for both the proteins, with axes fixed to $r_a$ = 1.8 nm and $r_b$ = 6.1 nm. 
The results for the normalized osmotic second virial coefficient for HSA obtained from our approach and from SAXS measurements are plotted in Fig.~\ref{fig:b2hsa}. Both experiments were performed under similar conditions and the overall agreement, despite the fact that in the different experiments $B_2$ is extracted from different data, is good. 

\begin{figure*}[!ht]
\includegraphics[scale=0.45]{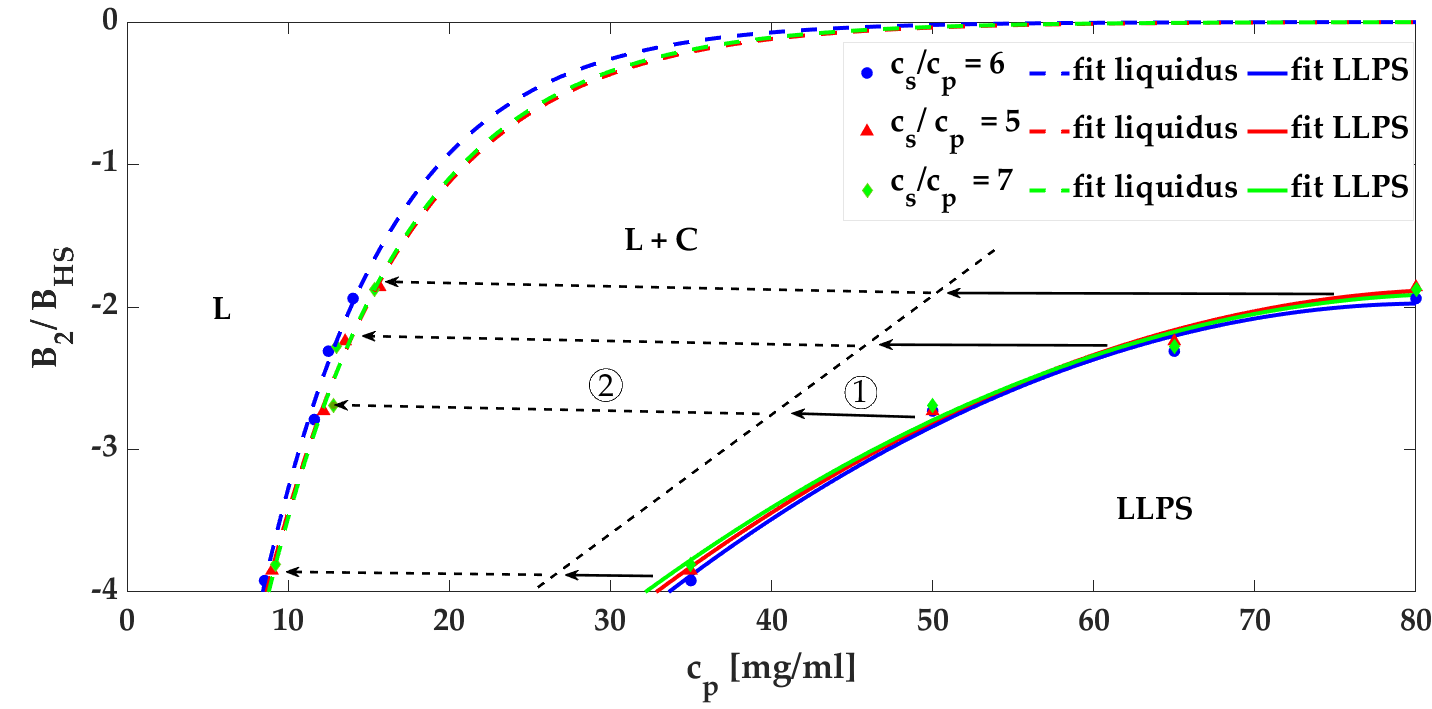}
\caption{\label{fig:phasehsa} Experimental phase diagram of HSA with CeCl$_3$. The position of the solubility line (liquidus) follows the magnitude of the interactions and the different partitioning effect at different protein salt-ratios. For $c_s \slash c_p$ = 6 the solubility of the protein diminish due to the stronger attractions. The overall mechanism of LLPS and crystallization of HSA in presence of CeCl$_3$ might be explained as follows: the system from the initial preparation conditions phase separate in a low and in a high density phase (1), from the low density phase the crystals start to nucleate and the consumption of material proceed until the solubility line is reached (2). The amount of high density and low density phase produced follow the arrows intensities.} \label{•}
\end{figure*}

\begin{table}[htb]
\centering
\resizebox{8cm}{!}{
\begin{tabular}{|l | c | c | c | r |}
\cline{1-5}
\multicolumn{5}{|c|} {HSA} \\ 
\hline
$c_p$[mg/ml] & $c_s$[mM]   & $c_s/c_p$  & $B'_2(supersaturation)$ & $\centering B'_2(SAXS)$ \qquad \quad \\
\hline
35 & 2.6  &  5 & -3.85 &  \\
35 & 3.2  &  6  & -3.92 &  \\
35 & 3.7  &  7  &  -3.81 & \\
\hline
50  &  3.8   & 5 &  -2.73  & -2.47 ($c_s/c_p$ = 5.3) \\
50  &  4.5   & 6 &  -2.79  & -2.72 ($c_s/c_p$ = 6.7) \\
50  &  5.2   & 7 &   -2.69 & -2.31 ($c_s/c_p$ = 8.0) \\
\hline
65 & 4.9  & 5 &  -2.24 & \\
65 & 5.9 & 6  &  -2.31 & \\
65 & 6.9 & 7  &  -2.28 & \\
\hline
80 & 6  & 5& -1.86 & \\
80 & 7.2 & 6 & -1.94 & \\
80 & 8.4 & 7 & -1.88 & \\
\hline
\end{tabular}
}
\caption{\label{tab:hsa} Normalized second virial coefficient values for HSA obtained via supersaturation. The results at 50 mg$\slash$ml are compared with those obtained from SAXS fitting. The error of the second virial coefficient obtained from supersaturation is about 10$\%$. $B'_2$. Values of $c_s/c_p$ are approximated by an integer number.}\end{table} 
For BSA the experimental conditions explored with SAXS were far from the LLPS border and additional analysis shall be performed for direct comparison of the $B'_2$ calculated with these two methods for the BSA-CeCl$_3$ system. However, a good agreement was found comparing the results obtained at 80 mg$\slash$ml from supersaturation measurements of BSA with CeCl$_3$ with those obtained with the SAXS fitting procedure at 80 mg$\slash$ml BSA concentration with LaCl$_3$ from ref.\cite{Matsarskaia_2018_PhysChemChemPhys} as shown in Tab.~\ref{tab:bsa}. Clearly, the effect of salt should produce differences in the magnitude of $B'_2$, due to effects such as differences in binding sites affinity on the surface of the protein or ions polarizability  that leads to different intermolecular interactions \cite{dumetz2007patterns, curtis2002protein}. We have chosen solutions with CeCl$_3$ for comparison, because these exhibit LLPS with to respect both HSA and BSA, while for LaCl$_3$ we obtain LLPS with HSA, but not for BSA. In BSA with LaCl$_3$ we observed transitions between Regime I (clear) – Regime II (turbidity without LLPS) and Regime II (turbidity without LLPS) - Regime III (reentrant). In principle we could induce partitioning in BSA with LaCl$_3$ by centrifuging the samples, while with CeCl$_3$ this is not necessary. Since our method relies only on thermodynamical assumptions, we apply the method exclusively on systems that show spontaneous driving force for LLPS. Independent of the type of protein used in this work, the general trend observed in both protein systems was an increasing of second virial coefficient as the protein concentration is increased, as shown in Tabs.~\ref{tab:hsa} and \ref{tab:bsa}. It is not clear if this behavior is directly derived from the mass transfer treatment discussed in Sec.~\ref{sec:th} or if it is because there are deviations from ideality at higher protein concentration and the approximation $\gamma_{eq}$ = 1 might not be valid even at the protein volume fraction of the low density equilibrium phase. However, this trend seems to describes the phase behavior of colloidal fluids near coexistence \cite{menon1991new, verduin1995phase, miller2004phase}. For HSA with CeCl$_3$ at this condition, the solubility line is located at $\approx$10 mg$\slash$ml and the LLPS border is around $\approx$ 30 mg$\slash$ml. In comparison, BSA solution is far more soluble than HSA and a smaller equilibrium concentration found was around 67 mg$\slash$ml. The values of the solubility line and LLPS borders depend of the initial protein concentration, salt concentration and their ratio.
Using the data obtained in Tab.~\ref{tab:hsa}, as well as the initial concentration of HSA and the equilibrium one, it was possible to rebuild an experimental and qualitative phase diagram that resembles the theoretical diagram for colloidal fluids Fig.~\ref{fig:phasehsa}. For the complete phase diagram additional samples at higher protein concentration could be investigated as this might provide additional information for the observed trend of the second virial coefficient. In fact, from the theory of colloidal fluids a turning point should be reached, corresponding to the critical point, and after that a decrease in $B'_2$ should be observed at higher protein volume fractions. However, this could not be detectable with the previous assumptions for the limitation due to the approximations made. Furthermore, for a complete description of the phase diagram, the concentration in the high density phase should be determined so that the border at higher protein concentrations of the LLPS can also be calculated. In Fig.~\ref{fig:phasehsa} a straight dotted line is used to describe the concentration of the low density phase after phase separation. This line could in principle be determined experimentally, if the sample is centrifuged at 5000 rpm for 10 minutes immediately after salt addition, to separate the contribution of LLPS (a kinetically driven process) to the contribution of the crystallization (a thermodynamically driven process).
\begin{table}[htb]
\centering
\resizebox{8cm}{!}{
\begin{tabular}{|l | c | c | c | r |}
\cline{1-5}
\multicolumn{5}{|c|} {BSA} \\
\hline
$c_p$[mg/ml] & $c_s$[mM]   & $c_s/c_p$  & $B'_2(supersaturation)$ & $\centering B'_2(SAXS)$ \qquad \quad \\
\hline
50 & 7 & 9 & NO & -2.41\\
50 & 8 & 11 & LLPS & -2.43\\
50 & 9 & 12 &  & -2.43\\
\hline
80 & 11 & 9 & -1.0 & \\
80 & 12 & 10 & -1.0 &\\
80 & 13 & 11 & -0.96 &\\
\hline
80 & 8 & 7 & & -1.0  (LaCl$_3$) \\
80 & 12 & 10 & &-1.5 (LaCl$_3$) \\
80 & 17 & 14 & &-1.2 (LaCl$_3$) \\
\hline
100 & 14 & 9 & -0.79 & \\
100 & 15 & 10 & -0.8  & \\
100 & 16 & 11 & -0.83 & \\
\hline
\end{tabular}
}
\caption{\label{tab:bsa} Normalized second virial coefficient values for BSA obtained via supersaturation. Note that no phase separation was observed at 50 mg$\slash$ml. The estimated error on the second virial coefficient is about 10$\%$. For 80 mg$\slash$ml the results are similar to those obtained  with LaCl$_3$ from ref. \cite{Matsarskaia_2018_PhysChemChemPhys}. Values of $c_s/c_p$ are approximated by an integer number.}
\end{table}

\subsection{Variation of the chemical potential near coexistence}
In order to take into account variations of the chemical potential and to extend the range of application of the derived formula outside phase coexistence, it is possible to normalize the previous relation, Eq.~(\ref{eq:b2}) with respect to the minimum concentration where LLPS is shown $\tilde{c}$:
\begin{equation} \label{eq:b2norm}
B_2 = - \frac{\ln\left(\frac{c_{eq}}{c_i}\right)}{2 M (c_{eq}-c_i)} \frac{c_i}{\tilde{c}}
\end{equation}
where $\tilde{c}$ is the lowest protein concentration for which LLPS is observed a given temperature and pressure. Taking into account this change in the chemical potential it is also possible to re-scale the second virial coefficient in such way that at constant supersaturation correspond always the same value of the second virial coefficient. Using this method, the concentration dependence vanishes. However, this is an empirical correction and additional studies are needed to explore the validity of the assumptions done.
The minimum concentration $\tilde{c}$ where LLPS is shown for HSA at salt-protein ratio of 6 is around 35-40 mg$\slash$ml and for BSA, at salt-protein ratio of 10 the minimum concentration $\tilde{c}$ is around 72-80 mg$\slash$ml. By using those values, the normalized second virial coefficient becomes, -3.4 for HSA at initial concentration of 35 and 50 mg$\slash$ml. The HSA system shows also an increased supersaturation by increasing protein concentration, by taking into account this behavior, the resulting second virial coefficient normalized for the minimum concentration where LLPS is shown $\tilde{c}$ at initial concentration of 65 and 80 mg$\slash$ml is approximately -3.7. For BSA the initial-equilibrium concentration ratio is approximately the same at each condition explored, around 1.2. Normalizing such system with respect to the critical concentration of 72  mg$\slash$, we obtain a normalized second virial coefficient of about -1.12 close to the theoretical limit for phase separation of sticky hard sphere if specific energies routes are followed \cite{wolf2014effective}. With this information, we speculate that for the BSA system we are moving close to the critical point in the phase diagram while for HSA we observed the optimal crystallization window below the critical point as predicted by G. A. Vliegenthart and H. N. W. Lekkerkerker \cite{doi:10.1063/1.481106}.

\section{Conclusions} \label{sec:con}
In conclusion, we calculated the normalized second virial coefficient for a protein solution close to LLPS using a thermodynamic approach for a two component system under the assumption that salt addition only changes the surface charge density of the proteins. Furthermore, we linked the macroscopic behavior of protein partitioning with the magnitude of the second virial coefficient that is a free model calculation and experimentally accessible. The magnitude of the second virial coefficient follows the partitioning ratio between two phases as previously observed for a different system \cite{wolf2014effective, Matsarskaia_2018_PhysChemChemPhys}. The mass transfer consideration might be in principle applied also outside the LLPS region, when the system reaches the solubility line. However, this strongly depends of the protein phase behavior, the molecular weight and the precipitation agent employed. Our expression was developed with the aim of being applicable to solutions containing two to three components, which exhibit both stable and metastable liquid-liquid phase separation (LLPS). This includes globular proteins with molecular weights ranging from 10 to 400 KDa, which covers, all the proteins investigated in reference \cite{dumetz2007patterns}. We have compared our results with those obtained from a different method finding a good agreement. The main constraint of the thermodynamical relation adopted is its 2-component approach. In fact, we access the chemical potential of the protein using the chemical potential of the solvent and the Gibbs-Duhem equation. As many solutions contain multiple components, such as buffers, salts, proteins, PEG (polyethylene glycol), water, preservatives, impurities and surfaces of glass vials or containers, the LLPS process and the supersaturation measurements may be influenced by the presence of those compounds, decreasing the accuracy on the estimation of the second virial coefficient. Another constraint stems from the prerequisite understanding of the phase diagram and accurate identification of the binodal region. Additionally, smaller negatively charged proteins may experience the effects of bridging, resulting in significant kinetic amorphous aggregation or a greater driving force for crystallization than observed in the current system. When the interactions between proteins are too strong, the dynamic aspect of the interactions plays a crucial role on determine the final state of the system. At this condition, Monte Carlo simulations of particles with hard cores and isotropic, square-well interactions, using the fluctuation-dissipation theorem, show that if no ordered bounds are present, like amorphous aggregation, the system do not relax in the more favourable thermodynamical state. Consequently, if not ordered cluster are formed the bonding breaking energy is too high that the system is confined in an energetic trap \cite{klotsa2011predicting}. In fact, the pathway for the formation of a new phase can be near equilibrium or far from it. The liquid-liquid phase separation process always competes with other forms of self-organization, including kinetic intermediates as previously discussed by Whitelam and Jack \cite{whitelam2015statistical}.
Under these conditions the treatment used could become inaccurate.  However, by taking into account the distance from the LLPS border, with the normalization used in Eq.~(\ref{eq:b2norm}) it might be possible to estimate the second virial coefficient also outside the binodal region. The accuracy of the treatment strongly decreases if after phase separation the concentration of the protein in the low density phase is still high, probably due to protein-protein interactions still present in this phase that should be also taken into account. In this work we have shown that under the conditions investigated the magnitude of the second virial coefficient calculated via supersaturation measurements is in agreement with the values obtained via SAXS fitting procedure. Working on the basic ideas presented here, might lead to a different approach to estimate the second virial coefficient that do not make use of expansive and advanced techniques. The method in fact could be employed routinely in any protein crystallization laboratories, without the need of training to perform complicated experiments. Furthermore, the method could be useful not only for proteins-salt solutions, but also for different colloidal systems that show crystallization or stable and metastable liquid-liquid phase separation.

\section{Acknowledgements}
The authors gratefully acknowledge the financial support from the DFG and the allocation of beamtime as well as the support of Madeleine Fries and Alexander Gerlach. They appreciate fruitful discussions with Anita Girelli. They thank Hadra Banks, Cara Buchholz and Lara Reichart for experimental assistance.

\bibliographystyle{acm}
\bibliography{An_Alternative}

\end{document}